%% file: jwfd.tex
\documentclass[universe,article,submit,moreauthors]{Definitions/mdpi}
\AtEndPreamble{\renewcommand{\linenumbers}{\relax}}

\usepackage{amsmath}
\usepackage{booktabs}
\usepackage{multirow}
\usepackage{tikz}
\usepackage[normalem]{ulem} 

\newif\ifrevision
\revisionfalse  

\newcommand{\add}[1]{\ifrevision\textcolor{blue}{#1}\else#1\fi}

\newcommand{\del}[1]{\ifrevision\textcolor{red}{\sout{#1}}\else\fi}

\newcommand{\chg}[2]{\del{#1}\add{#2}}
\usepackage{pgfplots}
\pgfplotsset{compat=1.18}
\usetikzlibrary{positioning,arrows.meta,shapes.geometric}

\firstpage{1}
\pubvolume{1}
\issuenum{1}
\articlenumber{0}
\pubyear{2026}
\copyrightyear{2026}
\datereceived{ }
\daterevised{ }
\dateaccepted{ }
\datepublished{ }

\Title{JW-FD\_V2: A 15-Year Multimodal Dataset for Solar Flare Forecasting}

\Author{Shao Mingfu $^{1}$\orcidA{}, Lin Jiaben $^{1,}$*, Wang Hui $^{1,2}$, Tong Liyue $^{1}$, Yang Chen $^{1,2}$, Zhang Yin $^{1,2}$ and Li Yuyang $^{2,1}$}

\AuthorNames{Shao Mingfu, Lin Jiaben, Wang Hui, Tong Liyue, Yang Chen, Zhang Yin, and Li Yuyang}

\address{%
$^{1}$ \quad State Key Laboratory of Solar Activity and Space Weather, NAOC, Beijing 100101, P.~R.~China; jiabenlin@bao.ac.cn\\
$^{2}$ \quad University of Chinese Academy of Sciences, Beijing 101408, P.~R.~China}

\corres{Correspondence: jiabenlin@bao.ac.cn}

\abstract{Solar flares drive severe space weather hazards, and forecasting their occurrence remains a central challenge for both heliophysics and operational space weather services. Data driven methods require long horizon datasets in which images, magnetic features, and flare labels are coregistered in space and time. We present \textbf{JW-FD} (JinWu-Flare Dataset), a 15 year multimodal release spanning 1~January~2011 through 31~December~2025, constructed from SDO/HMI line of sight magnetograms, NOAA Solar Region Summary reports, and NOAA X-ray flare event lists. The dataset comprises 3{,}083 independent active regions and 2{,}003{,}969 coregistered magnetogram crops, together with FITS, PNG, CSV, and MP4 modalities. Each sample provides 29 magnetic features linked to configurable flare labels under a strict pre-eruption window spanning seven forecast horizons and four GOES intensity thresholds. \chg{An 8:1:1 split at the active region level (random seed~62) keeps each active region in a single partition to prevent temporal leakage. Nineteen regions are then reassigned so that rare M- and X-class events are balanced between validation and test.}{An 8:1:1 split at the active region level, obtained by stratified rejection sampling (seed~3970), keeps each active region in a single partition to prevent temporal leakage and balances rare M- and X-class positives between validation and test.} PNG branches are released at ten magnetic saturation thresholds. \chg{Internal Transformer and four-model 24~h experiments on $\geq$C1.0 and $\geq$M1.0 forecasting suggest $B_{\mathrm{th}}=1000$~G as a preliminary default, although skill remains modest and the optimal saturation is model and task dependent.}{A Swin ablation over all ten clips finds $\geq$C1.0 skill nearly insensitive to $B_{\mathrm{th}}$, whereas $\geq$M1.0, $\geq$M5.0, and $\geq$X1.0 lock different clips. We therefore adopt $B_{\mathrm{th}}=800$~G as a compromise PNG working point rather than the optimum of any one task, and report four-model 24~h baselines at that clip. Skill remains modest as the class becomes rarer.} The open source construction pipeline is available at \url{https://github.com/Xiaoxuan-1/JW-FD}, and the released dataset (version~v2) at \url{https://doi.org/10.57760/sciencedb.019mg}.}


\begin{document}

\section{Introduction}
\label{sec:intro}

Solar flares are among the most energetic manifestations of solar magnetic activity. Major (M- and X-class) flares can disrupt satellites, communications, and power grids, making reliable flare forecasting a priority for both fundamental solar physics and operational space weather services~\cite{Shao2025review,Toriumi2019,Leka2019}.

Photospheric line of sight (LoS) magnetograms encode signatures of magnetic complexity (gradient
concentrations, neutral line topology, and flux imbalance) that precede many eruptive events~\cite{Boucheron2015,Bobra2014}. Quantitative features derived from magnetograms have long supported statistical and machine learning (ML) flare forecasting, and recent work extends this paradigm to convolutional neural networks (CNNs), vision transformers, and multimodal models that consume both images and structured parameters~\cite{Shao2025jwflare,Shao2026jwvl}. These methods require datasets in which images, physical features, and flare labels are coregistered in space and time, under a partitioning scheme that prevents temporal leakage across splits.

Several public datasets now support data driven flare forecasting, but they differ in temporal span, modality coverage, and evaluation design. Boucheron et al.\ released an AR LoS magnetogram image corpus covering 2010--2018~\cite{Boucheron2023}. Angryk et al.\ released SWAN-SF, a multivariate time series corpus of Space weather HMI Active Region Patch (SHARP) parameters over a similar interval~\cite{Angryk2020,Bobra2014sharp}. Nishizuka et al.\ assembled feature databases associated with Deep Flare Net (DeFN)~\cite{Nishizuka2017,Nishizuka2018}, and related work continues to organize SHARP parameter sets for operational forecasting~\cite{Abduallah2023,Florios2018}. Complementary studies have assembled magnetogram or multiwavelength datasets for CNNs, long short-term memory (LSTM) networks, and related models~\cite{Huang2018,Li2020,Sun2022,Jonas2018}, yet many remain study-specific and are not publicly redistributed as reusable, long-horizon, multimodal corpora~\cite{Shao2025review,Bloomfield2012,Barnes2016}.

More recently, SuryaBench provides a large multitask heliophysics benchmark for foundation model research, with solar flare forecasting as one of several applications~\cite{Roy2026}. Its flare subset uses full disk SDO context with a fixed 24~h labeling window and a calendar based train/test partition, rather than NOAA AR crops with configurable multi horizon labels and an AR level split. Relative to Boucheron et al.~\cite{Boucheron2023}, SWAN-SF~\cite{Angryk2020}, and the SuryaBench flare task~\cite{Roy2026}, the JinWu-Flare Dataset (JW-FD) supplies a long horizon, AR centered multimodal release in which images, magnetic features, flare labels, and evolution videos share a common index under a leakage aware AR level partition~\cite{Leka2019}.

\textbf{JW-FD} is designed to meet this need. This paper presents the dataset, its design rationale, construction pipeline, and recommended usage.
Our principal contributions are as follows:
\begin{enumerate}
\item A \textbf{15 year multimodal release} (2011--2025) covering solar Cycles~24--25 with 2.00~million active region magnetogram crops, supporting studies of cycle dependent flare statistics and rare X class events.
\item \textbf{Coregistered images, magnetic features, flare labels, and evolution videos}, with 66 dimensional records that link 29 magnetic features to configurable labels defined before eruption.
\item \textbf{Configurable labels} over seven forecast horizons (1--72~h) and four GOES intensity thresholds (C1.0--X1.0), defined on a strict pre-eruption window.
\item \textbf{An 8:1:1 training, validation, and test split.} The data are divided into training, validation, and test sets in an 8:1:1 ratio, with all observations from the same active region assigned to a single partition. This prevents information from the same region at different times from leaking between the datasets. \del{After the seed-62 random allocation, 19 active regions were reassigned so that rare M- and X-class positives are balanced between validation and test at all seven forecast horizons.}\add{The assignment is a stratified rejection sample (seed~3970) that equalizes rare-class positives between validation and test at all seven forecast horizons.}
\end{enumerate}

\section{Methods}
\label{sec:methods}

\begin{figure}[H]
\centering
\includegraphics[width=\textwidth]{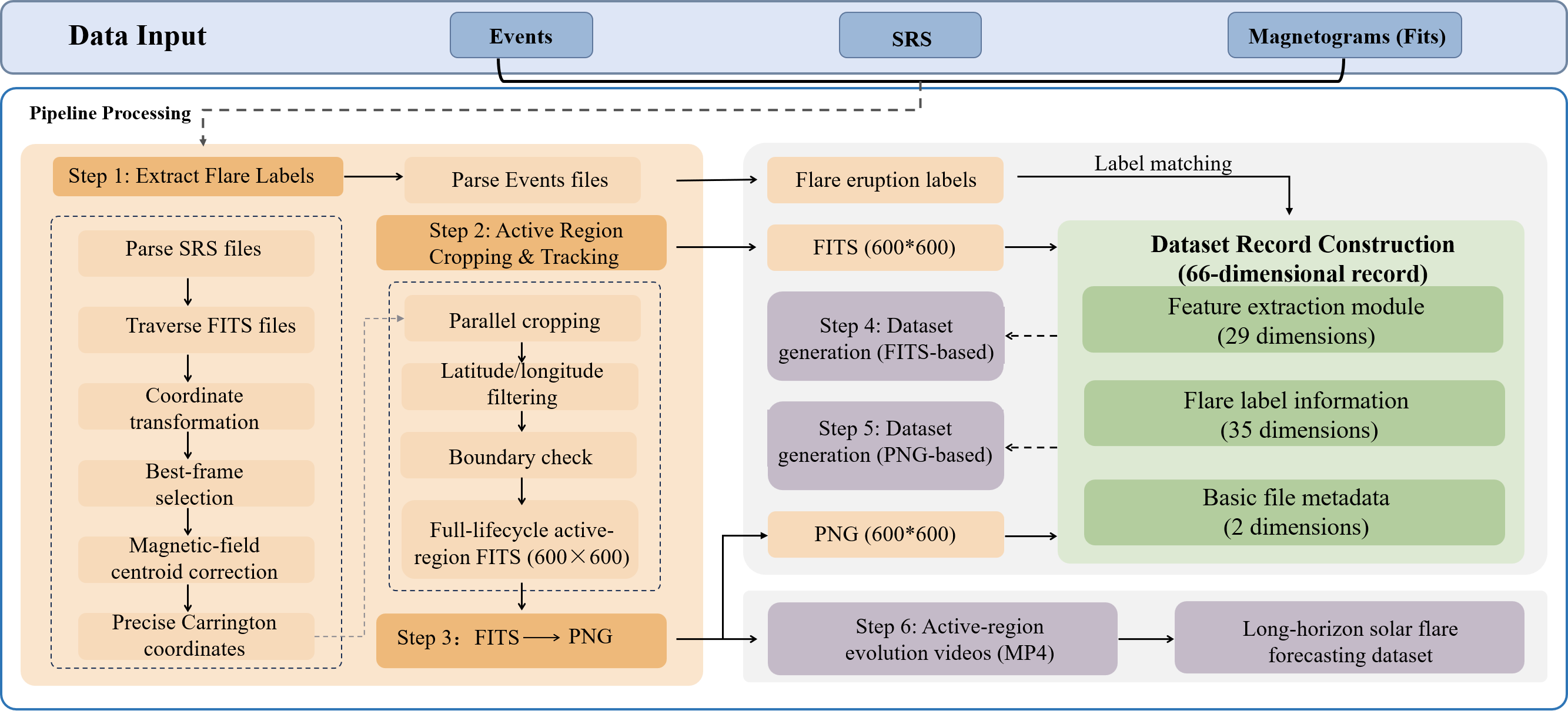}
\caption{JW-FD dataset construction workflow. Step~1 extracts NOAA X-ray flare events. Step~2 crops and tracks NOAA active regions on SDO/HMI full-disk line-of-sight magnetograms. Step~3 converts FITS crops to 8-bit PNG images at ten magnetic saturation thresholds. Steps~4 and~5 extract the same 29-dimensional feature set and share label-matching logic on the FITS and PNG branches, respectively. Step~6 encodes per-active-region PNG sequences as evolution videos.}
\label{fig:pipeline}
\end{figure}

JW-FD is constructed with a six step automated pipeline that ingests Solar Region Summary (SRS) reports and NOAA Events X-ray (XRA) flare lists from the National Oceanic and Atmospheric Administration/Space Weather Prediction Center (NOAA/SWPC), together with Solar Dynamics Observatory (SDO)/HMI~\cite{Scherrer2012,Pesnell2012} full disk LoS magnetograms. The pipeline exports Flexible Image Transport System (FITS) crops, Portable Network Graphics (PNG) images, comma separated values (CSV) tables, and MPEG-4 (MP4) videos. Figure~\ref{fig:pipeline} summarizes the workflow; Table~\ref{tab:pipeline} lists the inputs and outputs of each step.

\begin{table}[H]
\caption{Pipeline step inputs and outputs. Here $B_{\mathrm{th}}$ denotes the magnetic saturation threshold (in Gauss) applied when converting FITS magnetograms to 8-bit PNG images; the release includes ten branches from 200 to 2000~G in steps of 200~G.}
\label{tab:pipeline}
\centering
\scriptsize
\setlength{\tabcolsep}{4pt}
\begin{tabularx}{\textwidth}{@{}c>{\raggedright\arraybackslash}X>{\raggedright\arraybackslash}X>{\raggedright\arraybackslash}X@{}}
\toprule
\textbf{Step} & \textbf{Script} & \textbf{Input} & \textbf{Output} \\
\midrule
1 & \texttt{step1\_extract\_flare\_labels} & NOAA Events \texttt{.txt} (XRA) & \texttt{flare\_labels.csv} \\
2 & \texttt{step2\_crop\_active\_regions} & SRS + HMI FITS (4096$^2$) & \texttt{fits\_600/\{AR\}/} \\
3 & \texttt{step3\_convert\_to\_png} & Step~2 FITS & \texttt{png\_600\_Th\{$B_{\mathrm{th}}$\}/} \\
4 & \texttt{step4\_generate\_dataset\_fits} & FITS + labels & FITS branch CSV \\
5 & \texttt{step5\_generate\_dataset\_png} & PNG + labels & PNG branch CSV \\
6 & \texttt{step6\_make\_movie} & PNG sequences & \texttt{movies/AR\{num\}\_evolution.mp4} \\
\bottomrule
\end{tabularx}
\end{table}

\subsection{Data Sources}

\textbf{SDO/HMI} provides full disk LoS magnetograms at 4096$\times$4096 pixels ($\sim$1.5$''$ per pixel) with a cadence of approximately 12~minutes. The source frames \del{include limb-darkening correction but }are not projection-corrected: they remain in the observer plane, and $B_{\mathrm{LoS}}$ is not converted to an approximate radial field. \textbf{NOAA/SWPC SRS} reports list daily NOAA AR numbers and heliographic locations. \textbf{NOAA Events} files record GOES soft X-ray flare begin, maximum, and end times, together with GOES class strings (C, M, or X followed by a numeric subclass).

\subsection{Design Choices}
\label{sec:design}

Four design choices distinguish JW-FD from SHARP based patch datasets and earlier AR image datasets~\cite{Boucheron2023,Bobra2014}.

\textbf{Long temporal coverage.} Fifteen calendar years (2011--2025) span the declining phase of Solar Cycle~24, the 2018--2020 minimum, and the rising phase of Cycle~25. This span supports cycle dependent statistics and supplies enough rare M and X class events for threshold specific studies.

\textbf{Multimodal coregistration.} Every sample is indexed by NOAA AR number and Coordinated Universal Time (UTC) timestamp across FITS magnetograms (Gauss scale physics), PNG frames (vision model input), 29 dimensional magnetic feature vectors, configurable flare labels, and per AR MP4 evolution videos. Users may train end to end vision models on PNG or MP4 inputs, parameter based classifiers on extracted features, or multimodal models that fuse imagery with structured metadata~\cite{Shao2025jwflare}.

\textbf{Configurable pre-eruption labels.} Binary and class labels are defined for seven forecast horizons ($h \in \{1,3,6,12,24,48,72\}$~hours) and four GOES thresholds ($\tau \in \{\mathrm{C1.0,M1.0,M5.0,X1.0}\}$). Positive labels are assigned only within the half open interval $[\mathrm{Begin}-h,\,\mathrm{Begin})$, so frames at or after eruption onset are not treated as positive (Figure~\ref{fig:label_timeline}).

\textbf{AR level evaluation split.} Consecutive HMI frames of the same AR exhibit strong temporal autocorrelation, so assigning different timestamps of one AR to both training and test sets constitutes information leakage~\cite{Barnes2016,Leka2019}. JW-FD therefore adopts an 8:1:1 split at the AR level (Section~\ref{sec:validation})\del{; a subsequent 19-AR adjustment balances rare-class positives between validation and test}\add{, obtained by stratified rejection sampling so that rare-class positives are balanced between validation and test}.

Additional practical choices include: (i)~600$\times$600~pixel crops ($\sim$300$''$ field of view); (ii)~NOAA AR centered directory organization rather than HMI Active Region Patch (HARP)/SHARP patch identifiers, matching operational numbering; (iii)~a disk center filter ($|\mathrm{longitude}| \le 60^\circ$) that reduces the most severe limb effects but does not apply a geometric projection correction; and (iv)~dual FITS and PNG CSV branches so that physics faithful (Gauss scale) and vision ready (normalized intensity) feature extraction share identical label logic.

\input{figures/label_timeline.tex}

\subsection{Active Region Cropping and Tracking}

Tracking begins from the SRS Stonyhurst location of each NOAA AR at its first available magnetogram time, which is converted to a Carrington anchor. That anchor is then carried across subsequent epochs so that the same heliographic region remains centered as the Sun rotates. Local refinement masks pixels stronger than $100$~G inside an 80 pixel neighborhood of the projected center and updates the center to their mass centroid. From each full disk FITS frame we extract a $600\times600$ crop and discard ARs with $|\mathrm{longitude}| > 60^\circ$.

\subsection{FITS to PNG Conversion}

Cropped FITS magnetograms are converted to 8 bit PNG images for subsequent vision based analysis. Each crop is mirrored left to right to follow the conventional north up, east left display orientation. Magnetic field values are then clipped at a chosen saturation threshold and linearly mapped to the integer range $[0,255]$:
\begin{equation}
B' = \mathrm{clip}(B,\,-B_{\mathrm{th}},\,B_{\mathrm{th}}), \quad
I = \left\lfloor \frac{B' + B_{\mathrm{th}}}{2B_{\mathrm{th}}}\times 255 \right\rfloor,
\label{eq:norm}
\end{equation}
Invalid (NaN) samples are filled with $0$~G prior to clipping, so that $0$~G occupies grey level 127. We publish PNG trees for ten saturation levels, $B_{\mathrm{th}}=200$--$2000$~G in steps of $200$~G, each with matched 29-dimensional features and flare labels; Section~\ref{sec:threshold} discusses how this choice affects downstream models.

\subsection{Feature Extraction and Label Matching}
\label{sec:features}

Steps~4 and~5 generate CSV datasets from FITS and PNG inputs. Both branches extract the same 29 magnetic features (Table~\ref{tab:features}) and apply the same label-matching logic, but they operate on different physical domains. FITS features are computed on Gauss-scale magnetograms with no grey-level offset; invalid (NaN) pixels are replaced by $0$~G before extraction. PNG features are computed on 8-bit intensity images after subtracting a 127-level zero point, which is the uint8 bin that contains $0$~G under a fixed $\pm B_{\mathrm{th}}$ stretch (Equation~\ref{eq:norm}). The 29 features follow the snapshot (non-temporal) subset of the photospheric proxies introduced by Al-Ghraibah et al.~\cite{AlGhraibah2015} and subsequently used in a multivariate time-series setting by Boucheron et al.~\cite{Boucheron2015}. Al-Ghraibah et al.\ extracted flux, gradient, neutral-line, flux-evolution, and wavelet descriptors from line-of-sight magnetograms and classified whether an active region would produce a flare of GOES class $\geq$C1.0. JW-FD retains the 7 gradient, 13 neutral-line, 5 wavelet, and 4 flux descriptors (29 in total) as per-frame attributes; the 9 flux-evolution descriptors of Al-Ghraibah et al.\ are not precomputed, because the released image and video sequences already allow users to form temporal differences if desired. Operationally, (i)~Sobel filters yield a gradient-magnitude map from which we store the mean, standard deviation, median, extrema, skewness, and kurtosis; (ii)~a $10\times10$ boxcar mean (implemented as a uniform filter, with near-zero window means reset to $0$~G to avoid spurious contours from floating-point residue) and zero-Gauss contours define polarity inversion lines, which are then weighted by gradient magnitude to obtain length, fragment count, gradient-weighted length, and statistics of curvature and bending energy; (iii)~a five-level Haar decomposition yields orientation-summed sub-band energies L1--L5; and (iv)~total positive, negative, signed, and unsigned flux are obtained by summing pixel values over the corresponding polarity masks. These quantities are physically interpretable proxies of magnetic complexity and free-energy storage, and the open-source pipeline regenerates the same numbers from the released FITS crops.

\begin{figure}[H]
\centering
\includegraphics[width=\textwidth]{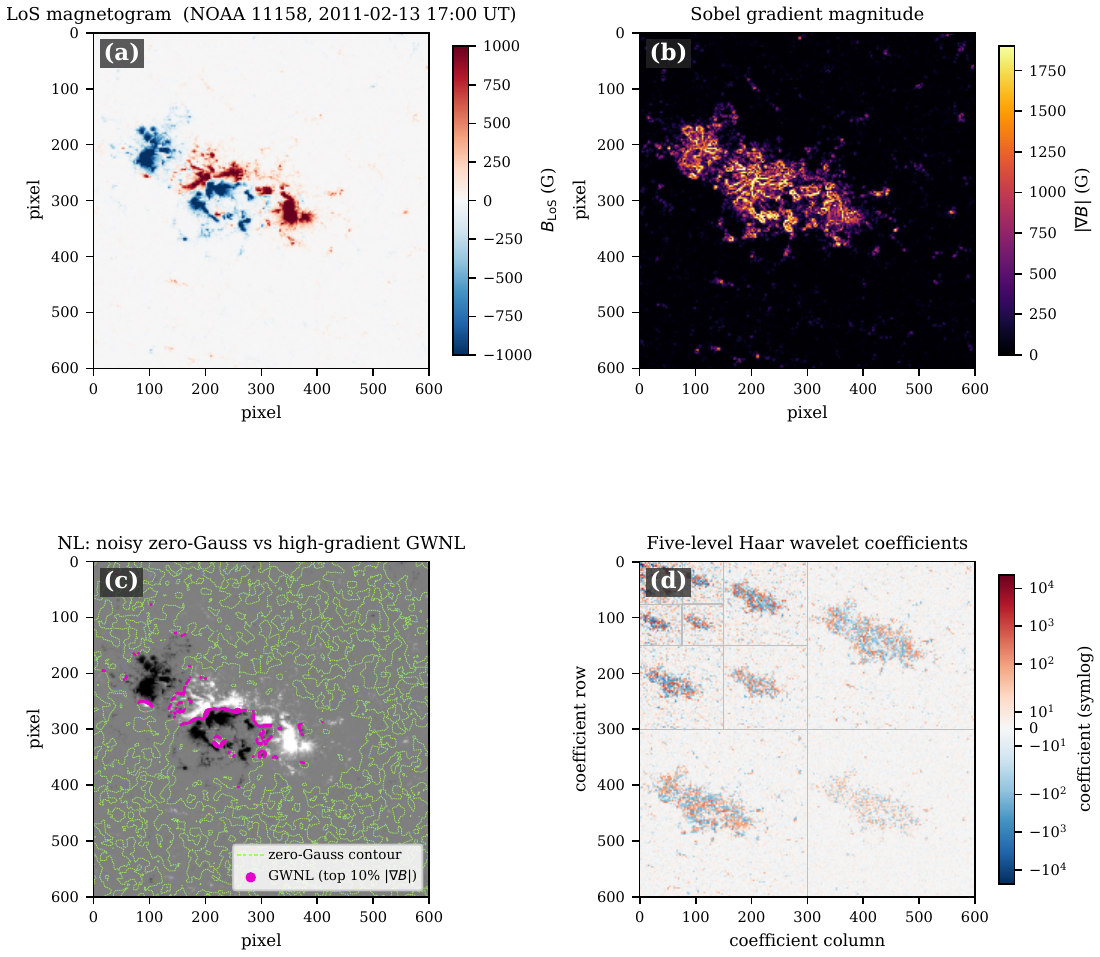}
\caption{Feature extraction on NOAA AR~11158 at 17:00~UT on 13~February~2011 (FITS crop, $600\times600$ pixels), following Al-Ghraibah et al.~\cite{AlGhraibah2015}. (a)~Line-of-sight magnetogram. (b)~Sobel gradient magnitude. (c)~Neutral-line contours and the gradient-weighted subset. (d)~Five-level Haar wavelet coefficients.}
\label{fig:feature_extraction}
\end{figure}

Figure~\ref{fig:feature_extraction} illustrates the four feature families on NOAA AR~11158 at 17:00~UT on 13~February~2011. Panel~(a) shows the line-of-sight magnetogram from which the four flux integrals are computed. Panel~(b) is the Sobel gradient-magnitude map that supplies the seven gradient statistics. Panel~(c) shows the neutral-line descriptors: thin lime dashes are all zero-Gauss contours after $10\times10$ smoothing (the noisy inversion line), and thin magenta markers mark only the high-gradient subset of the gradient-weighted neutral line (top 10\% of $|\nabla B|$ on the contour), with marker size scaled by $|\nabla B|$. Panel~(d) displays the five-level Haar coefficient array whose orientation-summed sub-band energies are stored as L1--L5.

\begin{table}[H]
\caption{Twenty-Nine Magnetic Features and Their CSV Column Names}
\label{tab:features}
\centering
\small
\begin{tabularx}{\textwidth}{@{}>{\raggedright\arraybackslash}p{2.8cm}X@{}}
\toprule
\textbf{Category} & \textbf{Feature names} \\
\midrule
Gradient (7) & Gradient mean; Gradient std; Gradient median; Gradient min; Gradient max; Gradient skewness; Gradient kurtosis \\
Neutral line (13) & NL length; NL no.\ fragments; NL gradient-weighted length; NL curvature mean/std/median/min/max; NL bending energy mean/std/median/min/max \\
Wavelet (5) & Wavelet Energy L1; Wavelet Energy L2; Wavelet Energy L3; Wavelet Energy L4; Wavelet Energy L5 \\
Flux (4) & Total positive flux; Total negative flux; Total signed flux; Total unsigned flux \\
\bottomrule
\end{tabularx}
\end{table}

Relative to image-only AR corpora~\cite{Boucheron2023}, SHARP parameter time series~\cite{Angryk2020}, and full-disk flare benchmarks~\cite{Roy2026}, JW-FD is distinctive here in that the 29 magnetic features, configurable flare labels, and FITS/PNG/MP4 products share a common AR-and-time index and can be regenerated from the released crops by the open-source pipeline.

Each CSV row corresponds to one magnetogram time stamp. We define seven forecast horizons $h \in \{1,3,6,12,24,48,72\}$~hours and four GOES thresholds $\tau \in \{\mathrm{C1.0,M1.0,M5.0,X1.0}\}$. The binary column \texttt{flare\_label\_\{$\tau$\}\_\{$h$hr\}} equals~1 when at least one flare in the same AR has GOES class $\ge\tau$ and a begin time $\mathrm{Begin}$ in the half open window
\begin{equation}
\mathrm{Begin} - h \;\le\; t \;<\; \mathrm{Begin}
\label{eq:window}
\end{equation}
and equals~0 otherwise. Separately, \texttt{flare\_class\_\{$h$hr\}} stores the strongest GOES class among flares that fall in the same window for horizon~$h$, and equals \texttt{0} when no such flare occurs.

\subsection{Evolution Videos}

For each AR directory, PNG frames are sorted chronologically and encoded as H.264 MP4 at 60~frames per second. Each AR produces one \texttt{AR\{num\}\_evolution.mp4} file in the \texttt{movies/} directory. The 60~fps setting is a playback rate for visualization, not an observational cadence: consecutive frames remain the native HMI LoS crops sampled at approximately 12~minutes. At this encoding, one second of video spans about 12~hours of solar time, so the MP4 files provide a time compressed overview of AR evolution for inspection and qualitative analysis rather than a high temporal resolution data stream.

\subsection{Quality Control and Expert Verification}
\label{sec:qc}

Quality control is enforced throughout the pipeline. Flare lists are first screened to remove records with invalid AR identifiers, missing peak times, or unparseable GOES class strings. During active region extraction, crops are retained only when their centers remain on the usable disk and the AR longitude lies within the prescribed limit. Subsequent conversion and feature extraction stages record truncated or unreadable inputs in step specific logs, skip empty files, and replace NaN values encountered during PNG normalization and FITS feature extraction (filled with $0$~G). These automated checks operate together with the feature pipeline in Section~\ref{sec:features} so that invalid flare records, unusable disk positions, and truncated or NaN inputs are excluded before the 29 magnetic parameters and flare labels are written.

In addition to these automated checks, solar physicists inspected a randomly selected 5\% subset of the processed samples. Inspection criteria included crop centering, completeness of the field of view, and consistency of M/X class labels with the intended windows before eruption. Samples affected by telemetry gaps or uncommon processing failures were excluded or regenerated.

\section{Results}
\label{sec:results}

\subsection{Dataset Records and Statistics}
\label{sec:records}

Table~\ref{tab:summary} summarizes the scale of the JW-FD release. Here each magnetogram sample is one $600\times600$ active region crop; the remaining rows inventory the accompanying modalities, flare event records, and evaluation split.

\begin{table}[H]
\caption{Summary statistics of the JW-FD release (2011--2025).}
\label{tab:summary}
\centering
\small
\begin{tabularx}{\textwidth}{@{}>{\raggedright\arraybackslash}p{4.5cm}X@{}}
\toprule
\textbf{Quantity} & \textbf{Specification} \\
\midrule
Temporal coverage & 1~January~2011 -- 31~December~2025 (15~yr) \\
Active regions & 3{,}083 independent NOAA ARs \\
Magnetogram samples & 2{,}003{,}969 \\
Spatial resolution & $600\times600$~pixels per AR crop \\
Released modalities & FITS, PNG, CSV, and MP4 (3{,}083 videos) \\
Flare event records & 18{,}986 NOAA XRA events (\texttt{flare\_labels.csv}) \\
Evaluation split & AR level 8:1:1 (training / validation / test) \\
Data sources & SDO/HMI LoS; NOAA/SWPC SRS; NOAA Events (XRA) \\
\bottomrule
\end{tabularx}
\end{table}

The 18{,}986 flare event records are the NOAA XRA events in 2011--2025 that carry an AR number (\texttt{flare\_labels.csv} contains 21{,}506 rows; the other 2{,}520 are events from 2010 or events without an AR number, and cannot be assigned to samples).

Annual sample totals are strongly modulated by the solar cycle (Figure~\ref{fig:year_chart}). Counts peak near the Cycle~24 maximum (2013--2014) and again during the Cycle~25 rise (2023--2025), whereas the 2018--2020 activity minimum yields far fewer NOAA numbered ARs and therefore fewer retained samples.

\input{figures/year_chart.tex}

Processed products share a common release root and are partitioned by modality (Table~\ref{tab:modalities}). Magnetogram and label directories are indexed by NOAA AR number; PNG branches further separate data by the saturation threshold $B_{\mathrm{th}}$.

\begin{table}[H]
\caption{Released data modalities and directory layout.}
\label{tab:modalities}
\centering
\small
\begin{tabularx}{\textwidth}{@{}>{\raggedright\arraybackslash}p{1.8cm}X>{\raggedright\arraybackslash}p{4.6cm}@{}}
\toprule
\textbf{Modality} & \textbf{Content} & \textbf{Directory layout} \\
\midrule
FITS & LoS magnetogram in Gauss; $600\times600$ crop & \texttt{fits\_600/\{ARnum\}/...} \\
PNG & 8 bit grayscale; saturation $B_{\mathrm{th}}$ encoded in path & \texttt{png\_600\_Th\{B\_th\}/\{ARnum\}/...} \\
CSV & 66 dimensional record per sample; FITS and PNG branches & \texttt{label/\{fits|png\}/Th\{B\_th\}/...} \\
MP4 & Per AR time compressed evolution (60~fps playback; $\sim$12~min source cadence) & \texttt{movies/AR\{num\}\_evolution.mp4} \\
Labels & NOAA XRA flare event list & \texttt{flare\_labels.csv} \\
\bottomrule
\end{tabularx}
\end{table}

Each CSV row indexes one sample under a fixed 66 dimensional schema (Table~\ref{tab:schema}). The same record can therefore support tabular learning, vision models, or multimodal fusion without reconstructing labels externally.

\begin{table}[H]
\caption{Composition of each 66 dimensional CSV record.}
\label{tab:schema}
\centering
\small
\begin{tabularx}{\textwidth}{@{}>{\raggedright\arraybackslash}p{2.8cm}c>{\raggedright\arraybackslash}X@{}}
\toprule
\textbf{Component} & \textbf{$N$} & \textbf{Description} \\
\midrule
Magnetic features & 29 & Gradient, neutral line, wavelet, and flux descriptors (Table~\ref{tab:features}) \\
Binary flare labels & 28 & Seven forecast horizons $\times$ four GOES thresholds \\
Class labels & 7 & Strongest GOES class within each forecast horizon \\
File metadata & 2 & \texttt{image\_filename} and \texttt{image\_path} \\
\midrule
\textbf{Total} & \textbf{66} & \\
\bottomrule
\end{tabularx}
\end{table}

For evaluation, JW-FD provides an AR-level 8:1:1 split (Table~\ref{tab:split}). \chg{The assignment begins as a random allocation of active regions with seed~62 and is then adjusted by moving 19 active regions so that rare M- and X-class events are not concentrated in validation or test.}{The assignment is a stratified rejection sample of active regions. Each NOAA region is placed in one flare-productivity stratum: multiple X-class, single X-class, M5 but not X, M1 but not M5, C-class only, or flare quiet. A stratum is divided into three tertiles by magnetogram-frame count when each tertile still contains at least 20 regions; smaller strata remain a single cell. Within each cell the regions are shuffled and cut 8:1:1, and every timestamp of a region follows that region into one partition. Candidate seeds are tried in order from 62, and a seed is discarded unless every gate passes. Validation and test must each contain at least five X-class and five M5-class positive regions. Separately for the $\geq$C1.0, $\geq$M1.0, $\geq$M5.0, and $\geq$X1.0 grades, the validation-to-test ratio of 24~h positive frames must lie in $[0.5,2]$, so that validation and test remain balanced. The flare-productivity strata equalize how many regions of each class fall in validation and test, but regions within one stratum can still differ in size and magnetic complexity. Each region is therefore summarized by two frame-wise medians, taken over all of its magnetograms: total unsigned flux, which measures how much magnetic flux the crop contains, and neutral-line length, which measures how extended the polarity boundary is. Cohen $d$ is the difference between the validation and test means of each median, divided by their pooled standard deviation, and is computed over every region in the two partitions. A seed is rejected when $|d|\ge 0.30$ for either quantity, that is, when one hold-out set is shifted by $0.3$ standard deviations or more toward larger or more complex regions. Because the decision threshold is selected on the validation set, this gate keeps that choice from being tuned on a magnetically easier or harder population than the test set. Seed~3970 is the first seed that meets all of these gates.} All samples from one AR are assigned to a single subset; sample fractions may depart slightly from 8:1:1 because ARs contribute unequal numbers of magnetograms. \del{C-class-only and flare-quiet regions were left on the seed-62 assignment. The 19 moves act on X-class, M5-but-not-X, and a few M1-but-not-M5 regions: the multi-X region NOAA~4274 is returned to training, four isolated-X regions with a complete 1--72~h positive-frame ladder (5, 15, 30, 60, 120, 240, and 360 frames at the native 12~min cadence) are placed in the test set, and validation and test are matched in the number of X-class (6/6) and M5-class (11/11) positive regions. The positive-frame table lists training, validation, and test counts at all seven forecast horizons. After the rebalance the validation-to-test ratios lie between 0.83 and 1.03 for X1.0 (previously 1.39--2.67) and between 0.84 and 1.00 for M5.0 (previously 2.17--2.81). The M1.0 ratio at 1~h is the farthest from unity (0.72), because that window has few positives; at 24--72~h the M1.0 ratios are 0.97--1.04. C1.0 ratios remain 0.87--0.97, as expected once C-only and quiet regions are held fixed.}\add{Validation and test each contain 307 regions. The 24~h positive regions are six X-class and 15 M5-class in validation, and six X-class and 13 M5-class in test. Two SWPC region numbers are corrected to the AIA 131~\AA\ bright core and the SSW Latest Events location: the X1.3 on 7 March 2012 belongs to NOAA~11429 rather than NOAA~11430, and the X1.1 on 23 March 2024 belongs to NOAA~13615 rather than NOAA~13614. Table~\ref{tab:split_pos} lists the training, validation, and test positive-frame counts at all seven forecast horizons.}

\begin{table}[H]
\caption{\chg{Active-region-level partitioning after the seed-62 random assignment and a 19-AR rare-class rebalance (nominal ratio 8:1:1).}{Active-region-level partitioning from stratified rejection sampling (seed~3970; nominal ratio 8:1:1).}}
\label{tab:split}
\centering
\small
\begin{tabular}{@{}lrrrr@{}}
\toprule
\textbf{Subset} & \textbf{Samples ($N$)} & \textbf{Fraction} & \textbf{ARs ($N$)} & \textbf{Fraction} \\
\midrule
Training & \chg{1{,}589{,}875}{1{,}604{,}907} & \chg{79.8\%}{80.09\%} & \chg{2{,}447}{2{,}469} & \chg{79.9\%}{80.08\%} \\
Validation & \chg{195{,}293}{202{,}019} & \chg{9.8\%}{10.08\%} & \chg{308}{307} & \chg{10.1\%}{9.96\%} \\
Test & \chg{206{,}079}{197{,}043} & \chg{10.3\%}{9.83\%} & \chg{309}{307} & \chg{10.1\%}{9.96\%} \\
\midrule
\textbf{Total} & \textbf{2{,}003{,}969} & \textbf{100\%} & \textbf{3{,}083} & \textbf{100\%} \\
\bottomrule
\end{tabular}
\end{table}

\begin{table}[!htbp]
\caption{\chg{Positive-frame counts in the training, validation, and test partitions after the 19-AR rebalance, for all seven forecast horizons and four GOES thresholds.}{Positive-frame counts in the training, validation, and test partitions under seed~3970, for all seven forecast horizons and four GOES thresholds.}}
\label{tab:split_pos}
\centering
\small
\setlength{\tabcolsep}{5pt}
\begin{tabular*}{\textwidth}{@{\extracolsep{\fill}}llrrrr@{}}
\toprule
\textbf{Horizon} & \textbf{Subset} & \textbf{$\geq$C1.0} & \textbf{$\geq$M1.0} & \textbf{$\geq$M5.0} & \textbf{$\geq$X1.0} \\
\midrule
\multirow{3}{*}{1~h} & Train & \chg{28{,}023}{43{,}601} & \chg{3{,}952}{5{,}786} & \chg{604}{884} & \chg{230}{335} \\
 & Val & \chg{3{,}512}{6{,}011} & \chg{590}{937} & \chg{106}{156} & \chg{30}{45} \\
 & Test & \chg{3{,}563}{5{,}482} & \chg{580}{759} & \chg{70}{90} & \chg{30}{40} \\
\midrule
\multirow{3}{*}{3~h} & Train & \chg{70{,}946}{109{,}216} & \chg{11{,}026}{15{,}978} & \chg{1{,}786}{2{,}595} & \chg{676}{975} \\
 & Val & \chg{9{,}112}{15{,}288} & \chg{1{,}666}{2{,}614} & \chg{314}{456} & \chg{86}{124} \\
 & Test & \chg{8{,}822}{13{,}351} & \chg{1{,}618}{2{,}107} & \chg{209}{269} & \chg{90}{120} \\
\midrule
\multirow{3}{*}{6~h} & Train & \chg{117{,}743}{179{,}286} & \chg{20{,}165}{29{,}025} & \chg{3{,}432}{5{,}031} & \chg{1{,}331}{1{,}921} \\
 & Val & \chg{15{,}157}{25{,}075} & \chg{3{,}001}{4{,}695} & \chg{602}{877} & \chg{161}{242} \\
 & Test & \chg{14{,}551}{21{,}614} & \chg{2{,}990}{3{,}844} & \chg{419}{539} & \chg{179}{239} \\
\midrule
\multirow{3}{*}{12~h} & Train & \chg{183{,}767}{275{,}093} & \chg{35{,}083}{49{,}737} & \chg{6{,}574}{9{,}672} & \chg{2{,}578}{3{,}748} \\
 & Val & \chg{24{,}037}{38{,}358} & \chg{5{,}138}{7{,}973} & \chg{1{,}094}{1{,}543} & \chg{307}{452} \\
 & Test & \chg{22{,}774}{32{,}817} & \chg{5{,}219}{6{,}609} & \chg{778}{1{,}009} & \chg{336}{447} \\
\midrule
\multirow{3}{*}{24~h} & Train & \chg{272{,}039}{399{,}097} & \chg{56{,}896}{79{,}471} & \chg{11{,}416}{16{,}867} & \chg{4{,}609}{6{,}746} \\
 & Val & \chg{35{,}259}{54{,}090} & \chg{8{,}377}{12{,}809} & \chg{1{,}963}{2{,}691} & \chg{554}{819} \\
 & Test & \chg{33{,}585}{47{,}676} & \chg{8{,}727}{11{,}113} & \chg{1{,}529}{1{,}946} & \chg{673}{850} \\
\midrule
\multirow{3}{*}{48~h} & Train & \chg{384{,}082}{546{,}922} & \chg{87{,}203}{119{,}917} & \chg{19{,}542}{28{,}466} & \chg{8{,}310}{11{,}989} \\
 & Val & \chg{48{,}565}{72{,}555} & \chg{12{,}987}{19{,}444} & \chg{3{,}228}{4{,}536} & \chg{1{,}081}{1{,}567} \\
 & Test & \chg{46{,}577}{64{,}779} & \chg{13{,}880}{17{,}615} & \chg{2{,}941}{3{,}807} & \chg{1{,}372}{1{,}660} \\
\midrule
\multirow{3}{*}{72~h} & Train & \chg{455{,}316}{634{,}729} & \chg{110{,}438}{151{,}606} & \chg{26{,}439}{37{,}937} & \chg{11{,}524}{16{,}535} \\
 & Val & \chg{56{,}583}{83{,}171} & \chg{16{,}343}{23{,}773} & \chg{4{,}357}{5{,}938} & \chg{1{,}672}{2{,}074} \\
 & Test & \chg{55{,}519}{74{,}796} & \chg{17{,}445}{21{,}698} & \chg{3{,}910}{5{,}358} & \chg{1{,}898}{2{,}298} \\
\bottomrule
\end{tabular*}
\end{table}

\subsection{Magnetic Saturation Threshold Ablation}
\label{sec:threshold}

Converting Gauss scale FITS magnetograms to 8 bit PNG requires a saturation level $B_{\mathrm{th}}$ in Equation~(\ref{eq:norm}). This choice affects visual contrast, feature statistics on PNG derived branches, and model input distributions. JW-FD therefore releases PNG branches at ten saturation levels, $B_{\mathrm{th}}=200$--$2000$~G in steps of $200$~G, each accompanied by 29-dimensional magnetic features and the same configurable flare labels.

To provide an initial reference point, we ran a controlled ablation over \chg{six of these values (200--1000~G and 2000~G) with a Transformer based vision classifier}{all ten values with a Swin-B vision classifier} on the PNG branch, using binary flare grades $\geq$C1.0, $\geq$M1.0, $\geq$M5.0, and $\geq$X1.0 and the \chg{seed-62 AR-level assignment, before the 19-AR rebalance}{seed~3970 AR-level assignment}. Thresholds are compared at a fixed decision point of $0.5$ on the validation set (val TSS@$0.5$), so that clip differences are not confounded by per-threshold tuning. Table~\ref{tab:threshold} reports that validation score together with the corresponding test TSS@$0.5$, ROC-AUC, and PR-AUC. Across the ten clips, val TSS@$0.5$ for $\geq$C1.0 spans only $0.5130$--$0.5371$, so $\geq$C1.0 skill is nearly insensitive to $B_{\mathrm{th}}$. \del{In this internal study, 1000~G attains the highest TPR (0.6320), TSS (0.5225), and MCC (0.2107), whereas 200~G is stronger on ACC, TNR, F1, Precision, and FAR.}

\begin{table}[!htbp]
\caption{Forecast performance under different magnetic saturation thresholds $B_{\mathrm{th}}$ \chg{(Transformer, C1.0, seed-62 AR-level split). For FAR, lower is better; for all other metrics, higher is better. Best values in each column are shown in bold.}{(Swin-B, 24~h, seed~3970) for $\geq$C1.0, $\geq$M1.0, $\geq$M5.0, and $\geq$X1.0. Clips are compared at a fixed decision threshold of $0.5$. The compromise working point $800$~G is shown in bold in each task block. The $\geq$X1.0 block is reported for completeness and is not used to select the release default.}}
\label{tab:threshold}
\centering
\small
\setlength{\tabcolsep}{5pt}
\begin{tabular*}{\textwidth}{@{\extracolsep{\fill}}llcccc@{}}
\toprule
\textbf{Task} & \textbf{$B_{\mathrm{th}}$} & \textbf{val TSS@$0.5$} & \textbf{test TSS@$0.5$} & \textbf{test ROC} & \textbf{test PR} \\
\midrule
\multirow{10}{*}{$\geq$C1.0} & 200~G & 0.5130 & 0.5078 & 0.8507 & 0.6110 \\
 & 400~G & 0.5242 & 0.5317 & 0.8487 & 0.6078 \\
 & 600~G & 0.5256 & 0.5251 & 0.8566 & 0.6122 \\
 & \textbf{800~G} & \textbf{0.5298} & \textbf{0.5282} & \textbf{0.8573} & \textbf{0.6205} \\
 & 1000~G & 0.5363 & 0.5327 & 0.8576 & 0.6207 \\
 & 1200~G & 0.5364 & 0.5266 & 0.8588 & 0.6238 \\
 & 1400~G & 0.5317 & 0.5440 & 0.8639 & 0.6239 \\
 & 1600~G & 0.5328 & 0.5326 & 0.8599 & 0.6222 \\
 & 1800~G & 0.5371 & 0.5317 & 0.8590 & 0.6161 \\
 & 2000~G & 0.5326 & 0.5196 & 0.8579 & 0.6142 \\
\midrule
\multirow{10}{*}{$\geq$M1.0} & 200~G & 0.6698 & 0.6396 & 0.9128 & 0.3625 \\
 & 400~G & 0.6799 & 0.6439 & 0.9161 & 0.3661 \\
 & 600~G & 0.6404 & 0.6501 & 0.9146 & 0.3896 \\
 & \textbf{800~G} & \textbf{0.6564} & \textbf{0.6746} & \textbf{0.9178} & \textbf{0.3758} \\
 & 1000~G & 0.6166 & 0.6058 & 0.9178 & 0.3922 \\
 & 1200~G & 0.6135 & 0.6147 & 0.9203 & 0.4065 \\
 & 1400~G & 0.6089 & 0.6230 & 0.9229 & 0.4207 \\
 & 1600~G & 0.6244 & 0.6397 & 0.9248 & 0.4250 \\
 & 1800~G & 0.6030 & 0.5951 & 0.9249 & 0.4211 \\
 & 2000~G & 0.6140 & 0.6042 & 0.9256 & 0.4189 \\
\midrule
\multirow{10}{*}{$\geq$M5.0} & 200~G & 0.7344 & 0.7938 & 0.9531 & 0.1020 \\
 & 400~G & 0.7300 & 0.8083 & 0.9548 & 0.1165 \\
 & 600~G & 0.7413 & 0.8151 & 0.9544 & 0.1207 \\
 & \textbf{800~G} & \textbf{0.7675} & \textbf{0.7761} & \textbf{0.9501} & \textbf{0.1153} \\
 & 1000~G & 0.7588 & 0.7300 & 0.9461 & 0.1118 \\
 & 1200~G & 0.7500 & 0.7248 & 0.9437 & 0.1085 \\
 & 1400~G & 0.7483 & 0.7323 & 0.9435 & 0.1066 \\
 & 1600~G & 0.7589 & 0.7316 & 0.9461 & 0.1104 \\
 & 1800~G & 0.7681 & 0.7292 & 0.9482 & 0.1077 \\
 & 2000~G & 0.7704 & 0.7145 & 0.9475 & 0.1037 \\
\midrule
\multirow{10}{*}{$\geq$X1.0} & 200~G & 0.6743 & 0.8790 & 0.9602 & 0.0459 \\
 & 400~G & 0.7422 & 0.8896 & 0.9616 & 0.0470 \\
 & 600~G & 0.7548 & 0.8774 & 0.9612 & 0.0464 \\
 & \textbf{800~G} & \textbf{0.7255} & \textbf{0.8892} & \textbf{0.9609} & \textbf{0.0463} \\
 & 1000~G & 0.7080 & 0.8852 & 0.9608 & 0.0478 \\
 & 1200~G & 0.6858 & 0.8688 & 0.9572 & 0.0448 \\
 & 1400~G & 0.6711 & 0.8285 & 0.9549 & 0.0433 \\
 & 1600~G & 0.6672 & 0.7905 & 0.9535 & 0.0426 \\
 & 1800~G & 0.6487 & 0.7394 & 0.9522 & 0.0422 \\
 & 2000~G & 0.6466 & 0.7397 & 0.9512 & 0.0419 \\
\bottomrule
\end{tabular*}
\end{table}

\add{Unlike $\geq$C1.0, the rarer-class tasks are more sensitive to $B_{\mathrm{th}}$ and do not share a val TSS@$0.5$ maximum: the lock is $400$~G for $\geq$M1.0 (range $0.6030$--$0.6799$), $2000$~G for $\geq$M5.0 ($0.7300$--$0.7704$), and $600$~G for $\geq$X1.0 ($0.6466$--$0.7548$). A monotonic trend in which stronger flares simply prefer higher clips is therefore not supported.}

\add{A low $B_{\mathrm{th}}$ saturates $|B|>B_{\mathrm{th}}$ and allocates grey levels to weak field and polarity-inversion structure; a high $B_{\mathrm{th}}$ preserves unsaturated umbral and penumbral field. The $\geq$C1.0 plateau is consistent with either cue being sufficient. The $\geq$M1.0 preference for $200$--$400$~G is compatible with inversion-line contrast, but this experiment uses PNG images rather than an ablation of the 29 magnetic-feature families, so that reading remains an interpretation. On the $\geq$M5.0 validation set the high clips $800$--$2000$~G form a cluster ($800$~G lies within $0.003$ of $2000$~G), which would fit a strong-field requirement. The $\geq$M5.0 test set does not confirm that pattern: test TSS@$0.5$, TSS at the validation-tuned operating point, ROC-AUC, and PR-AUC all favour mid-to-low clips (near $600$~G), and $2000$~G is among the weakest test scores. The $\geq$X1.0 grid is still more sample-limited and is not used to choose the working point.}

\add{The $800$~G clip is first in none of the four task blocks, yet it is never among the weakest. We therefore adopt $B_{\mathrm{th}}=800$~G as a single compromise working point for the released PNG branch, not as the optimal clip of any one task.}

We emphasize that this threshold selection is a preliminary exploration limited to the Swin architecture and a snapshot 24~h setting. Because JW-FD releases PNG branches at all ten $B_{\mathrm{th}}$ values, users should treat \chg{1000~G}{the $800$~G compromise} as a starting point rather than a universal optimum and re-ablate when the model, horizon, or GOES threshold differs. \del{That first ablation also omitted 1200, 1400, 1600, and 1800~G. After the 19-AR rebalance, both this ten-threshold grid and the four-model 24~h baselines need to be retrained on the released assignment: four isolated-X regions that were in the seed-62 training set now sit in the test set, so the present checkpoints cannot be scored on the released split.}

\subsection{Baseline Forecasting Experiments}
\label{sec:baseline}

To illustrate how JW-FD can be used for both parameter-based and vision-based forecasting, we trained four baselines on the \chg{seed-62 AR-level assignment}{seed~3970 AR-level assignment} at the compromise clip $B_{\mathrm{th}}={}$\chg{1000}{800}~G: an RBF support vector machine on the 29 magnetic features, and ResNet-152, ViT-B/16, and Swin-B classifiers on the PNG crops. These experiments are usage demonstrations of the released records at that shared clip, not a further per-model threshold search. The task is binary 24~h forecasting at GOES thresholds $\geq$C1.0, $\geq$M1.0, $\geq$M5.0, and $\geq$X1.0. Training used class-balanced losses; all active regions were retained, while magnetogram frames were temporally subsampled. Testing used the full \chg{204{,}920-frame seed-62 test partition}{test partition} at natural class prevalence. The decision threshold was selected on the validation set to maximize TSS. \del{These scores are an internal reference on the initial random split. They will be replaced once the same four models are retrained on the released 19-AR assignment (test 206,079 frames).}

\begin{figure}[!htbp]
\centering
\includegraphics[width=\textwidth]{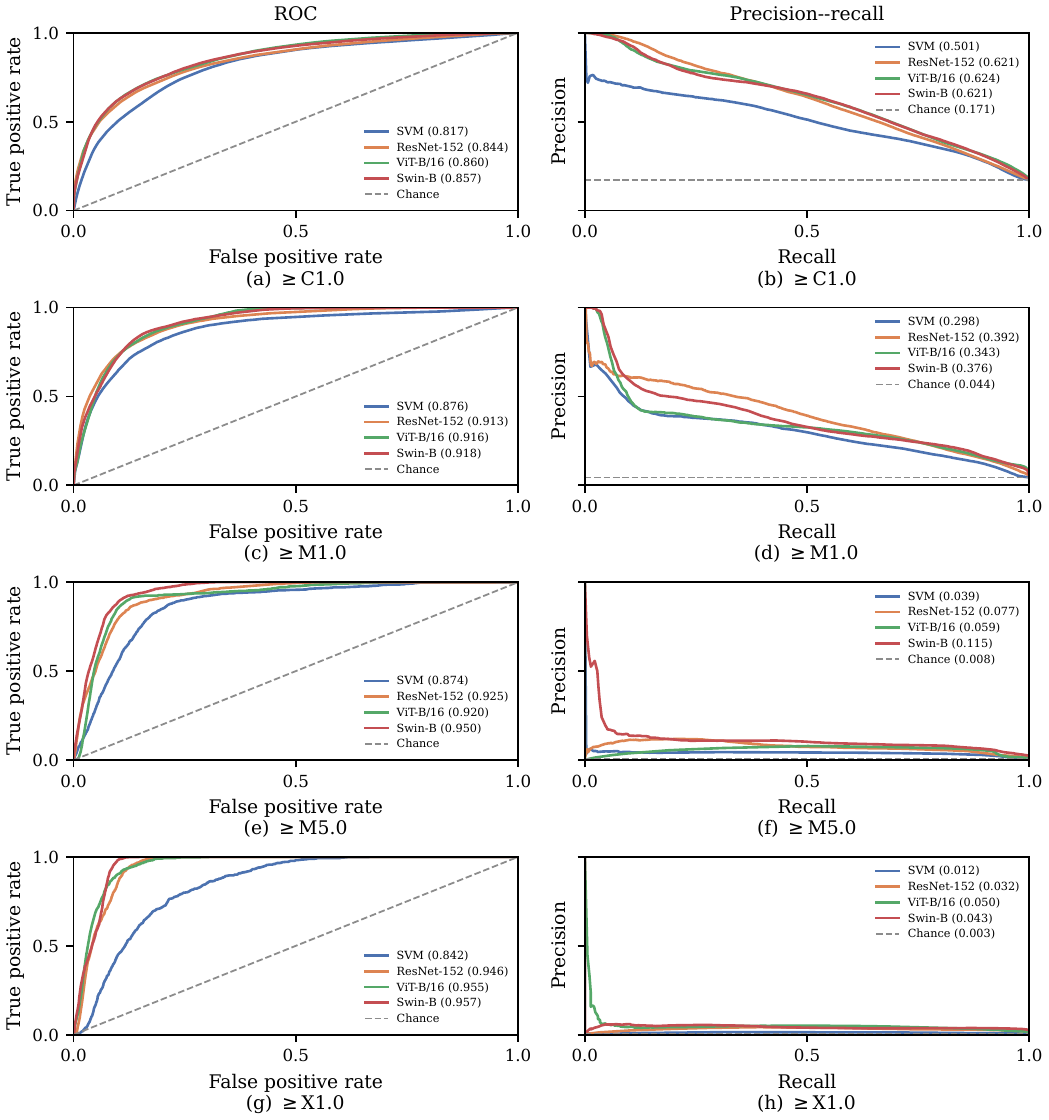}
\caption{\chg{ROC (top) and precision--recall (bottom) curves on the seed-62 AR-level test set for 24~h forecasting at C1.0 (left) and M1.0 (right).}{ROC (left) and precision--recall (right) curves on the seed~3970 AR-level test set for 24~h forecasting. Panels (a,b)~$\geq$C1.0; (c,d)~$\geq$M1.0; (e,f)~$\geq$M5.0; (g,h)~$\geq$X1.0. Chance is the diagonal on the ROC panels and the positive-class prevalence on the PR panels. Ranking remains high while precision collapses as the class becomes rarer.}}
\label{fig:roc_pr}
\end{figure}

Table~\ref{tab:baseline} reports test-set skill. For $\geq$C1.0, the three vision models attain TSS of order $0.52$--$0.55$ and F1 of order $0.57$--$0.58$, while SVM is slightly lower (TSS~$0.49$). For $\geq$M1.0, TSS rises to about $0.61$--$0.70$, while precision and F1 are lower because positives are rarer. The same pattern continues at $\geq$M5.0 and $\geq$X1.0: TSS remains high ($0.65$--$0.84$ for the vision models), but precision, F1, and PR-AUC collapse. The area under the receiver operating characteristic curve (ROC-AUC) and the area under the precision--recall curve (PR-AUC) do not depend on the TSS-tuned decision threshold. ROC-AUC asks only whether flare frames receive higher scores than non-flare frames; it remains $0.82$--$0.86$ for $\geq$C1.0 and $0.88$--$0.92$ for $\geq$M1.0, and stays above $0.84$ even for $\geq$X1.0. PR-AUC asks how often a high score is a true flare, and therefore tracks prevalence: it falls from $0.50$--$0.62$ at $\geq$C1.0 to $0.30$--$0.39$ at $\geq$M1.0, then to $0.04$--$0.12$ at $\geq$M5.0 and $0.01$--$0.05$ at $\geq$X1.0, in line with the FAR above $0.78$ for $\geq$M1.0 and above $0.92$ for $\geq$M5.0. \add{Figure~\ref{fig:roc_pr} shows the corresponding ROC and precision--recall curves.} These C- and M-class scores are comparable in magnitude to earlier LoS-parameter studies~\cite{AlGhraibah2015}. Dedicated SHARP-based systems such as FlareCast~\cite{Lv2026} report higher test-set skill, and two points help to place that comparison in context. First, FlareCast takes as input sequences of SHARP radial-field ($B_r$) magnetograms (six frames spanning 6~h), rather than line-of-sight PNG crops, and it first regresses the maximum X-ray flux before assigning a flare class. Second, the 24~h test-set TSS of 0.891 for $\geq$C is accompanied by lower values for $\geq$M~(0.792) and $\geq$X~(0.831); in an operational window from 1~August to 24~November~2024, the four-class confusion-matrix diagonals are 0.58~(N), 0.57~(C), 0.63~(M), and 0.81~(X). These operational figures provide additional perspective on deployed skill, and the C-class diagonal of 0.57 is comparable to our $\geq$C1.0 test baselines.

\begin{table}[H]
\caption{\chg{Baseline 24~h forecast skill on the seed-62 AR-level test set (204{,}920 frames, before the 19-AR rebalance; 1000~G).}{Baseline 24~h forecast skill on the seed~3970 AR-level test set ($B_{\mathrm{th}}=800$~G).} Vision models use PNG crops; SVM uses the 29 magnetic features. Decision thresholds are selected on the validation set to maximize TSS. ROC and PR denote ROC-AUC and PR-AUC (threshold-free). For FAR, lower is better; for all other metrics, higher is better. Best values in each task block are shown in bold.}
\label{tab:baseline}
\centering
\small
\setlength{\tabcolsep}{2.6pt}
\begin{tabular*}{\textwidth}{@{\extracolsep{\fill}}llcccccccccc@{}}
\toprule
\textbf{Task} & \textbf{Model} & \textbf{ACC} & \textbf{TPR} & \textbf{TNR} & \textbf{TSS} & \textbf{F1} & \textbf{MCC} & \textbf{Prec.} & \textbf{FAR} & \textbf{ROC} & \textbf{PR} \\
\midrule
\multirow{4}{*}{$\geq$C1.0} & SVM & 0.7771 & 0.6910 & 0.7948 & 0.4859 & 0.5141 & 0.4036 & 0.4093 & 0.5907 & 0.8166 & 0.5010 \\
 & ResNet-152 & 0.8289 & 0.6608 & \textbf{0.8634} & 0.5243 & 0.5685 & 0.4715 & 0.4988 & 0.5012 & 0.8436 & 0.6208 \\
 & ViT-B/16 & 0.8191 & \textbf{0.7096} & 0.8417 & \textbf{0.5513} & 0.5724 & 0.4774 & 0.4797 & 0.5203 & \textbf{0.8601} & \textbf{0.6236} \\
 & Swin-B & \textbf{0.8312} & 0.6824 & 0.8618 & 0.5442 & \textbf{0.5797} & \textbf{0.4856} & \textbf{0.5039} & \textbf{0.4961} & 0.8573 & 0.6205 \\
\midrule
\multirow{4}{*}{$\geq$M1.0} & SVM & 0.8235 & 0.7873 & 0.8252 & 0.6125 & 0.2834 & 0.3140 & 0.1728 & 0.8272 & 0.8757 & 0.2981 \\
 & ResNet-152 & \textbf{0.8599} & 0.7983 & \textbf{0.8628} & 0.6611 & \textbf{0.3357} & \textbf{0.3653} & \textbf{0.2125} & \textbf{0.7875} & 0.9127 & \textbf{0.3923} \\
 & ViT-B/16 & 0.8134 & \textbf{0.8693} & 0.8109 & 0.6801 & 0.2924 & 0.3383 & 0.1757 & 0.8243 & 0.9159 & 0.3428 \\
 & Swin-B & 0.8297 & 0.8681 & 0.8279 & \textbf{0.6960} & 0.3113 & 0.3562 & 0.1897 & 0.8103 & \textbf{0.9178} & 0.3758 \\
\midrule
\multirow{4}{*}{$\geq$M5.0} & SVM & 0.7937 & 0.8613 & 0.7931 & 0.6545 & 0.0609 & 0.1406 & 0.0316 & 0.9684 & 0.8735 & 0.0390 \\
 & ResNet-152 & 0.8817 & 0.8411 & 0.8821 & 0.7231 & 0.0995 & 0.1929 & 0.0529 & 0.9471 & 0.9250 & 0.0771 \\
 & ViT-B/16 & 0.8254 & \textbf{0.9248} & 0.8246 & 0.7494 & 0.0760 & 0.1708 & 0.0396 & 0.9604 & 0.9201 & 0.0590 \\
 & Swin-B & \textbf{0.9118} & 0.8568 & \textbf{0.9123} & \textbf{0.7690} & \textbf{0.1312} & \textbf{0.2317} & \textbf{0.0710} & \textbf{0.9290} & \textbf{0.9501} & \textbf{0.1153} \\
\midrule
\multirow{4}{*}{$\geq$X1.0} & SVM & 0.6862 & 0.8648 & 0.6856 & 0.5504 & 0.0185 & 0.0691 & 0.0093 & 0.9907 & 0.8425 & 0.0117 \\
 & ResNet-152 & 0.8247 & 0.9896 & 0.8241 & 0.8137 & 0.0372 & 0.1240 & 0.0189 & 0.9811 & 0.9460 & 0.0322 \\
 & ViT-B/16 & \textbf{0.8770} & 0.9361 & \textbf{0.8768} & 0.8129 & \textbf{0.0495} & \textbf{0.1430} & \textbf{0.0254} & \textbf{0.9746} & 0.9548 & \textbf{0.0499} \\
 & Swin-B & 0.8431 & \textbf{1.0000} & 0.8426 & \textbf{0.8426} & 0.0418 & 0.1341 & 0.0213 & 0.9787 & \textbf{0.9575} & 0.0434 \\
\bottomrule
\end{tabular*}
\end{table}

\section{Discussion}
\label{sec:discussion}

\subsection{Evaluation Protocol and Class Imbalance}
\label{sec:validation}

JW-FD adopts an \textbf{8:1:1 split at the AR level} obtained by stratified rejection sampling (seed~3970), ensuring that all magnetogram crops from a given AR appear exclusively in the training, validation, or test partition (Table~\ref{tab:split}). This protocol prevents information leakage arising from temporal autocorrelation within ARs~\cite{Barnes2016,Leka2019}. \del{A random allocation by AR count does not, however, balance rare M5 and X positives: under seed~62 the X1.0 validation set contained 833 positive 24~h frames against 370 in test, largely because NOAA~4274 (five X-class flares) occupied validation. The released assignment therefore keeps the seed-62 membership for C-only and quiet regions and moves 19 flare-productive regions, matching validation and test in X-class (6/6) and M5-class (11/11) positive ARs. The positive-frame table reports the resulting counts in training, validation, and test at all seven horizons.}\add{The pre-registered gates require matched rare-class counts and bounded positive-frame ratios between validation and test; Table~\ref{tab:split_pos} reports the resulting positive-frame counts in training, validation, and test at all seven horizons.} Split CSV files (\texttt{*\_train.csv}, \texttt{*\_val.csv}, \texttt{*\_test.csv}) are distributed under \texttt{label/png/Th\{threshold\}/} and \texttt{label/fits/}.

Positive labels are rare for higher flare thresholds (M5.0, X1.0), producing pronounced class imbalance. At longer forecast horizons the positive fraction increases, because a qualifying flare anywhere in a longer future window yields a positive label; shorter horizons are therefore more imbalanced. Over the full release, sample-level GOES class labels at the 72~h horizon comprise 20{,}907~X-class, 176{,}735~M-class, and 595{,}054~C-class events, together with 1{,}211{,}273~non-flare samples. Skill oriented metrics such as TSS, F1, and MCC therefore provide a more balanced assessment than raw accuracy alone~\cite{Bloomfield2012}.

\subsection{Recommended Usage}
\label{sec:usage}

JW-FD is intended for the heliophysics and space weather ML communities. Suggested applications include:

\begin{itemize}
\item \textbf{Vision and video models}: Train CNNs, ViTs, or temporal models on PNG crops or MP4 evolution sequences, selecting $B_{\mathrm{th}}$ by ablation (Section~\ref{sec:threshold}) and comparing against the 24~h baselines in Section~\ref{sec:baseline}. \del{Until those baselines are retrained on the released 19-AR assignment, the baseline table should be read as a seed-62 reference rather than a score on the distributed test CSV.}\add{The compromise starting clip is $B_{\mathrm{th}}=800$~G.}
\item \textbf{Parameter-based ML}: Apply classical classifiers or gradient boosted trees to the 29 dimensional magnetic parameter vectors, comparing FITS derived (Gauss scale) and PNG derived (intensity scale) branches.
\item \textbf{Multimodal fusion}: Combine PNG/MP4 inputs with magnetic parameters and labels from the same indexed record~\cite{Shao2025jwflare,Shao2026jwvl}.
\item \textbf{Multi horizon benchmarking}: Evaluate models across seven forecast horizons and four GOES thresholds using the precomputed label columns.
\end{itemize}

The approximately 2.00 million magnetogram frames originate from 3{,}083 active regions and are strongly autocorrelated in time within each region; they should not be treated as two million independent physical samples. The 24~h baselines in Section~\ref{sec:baseline} make this concrete for snapshot models. Training retained every active region but temporally subsampled frames, and on the \chg{seed-62 AR-level test set}{seed~3970 AR-level test set} an SVM on the 29 magnetic features is only modestly below ResNet-152, ViT-B/16, and Swin-B. JW-FD therefore supports parameter-based ML and single-frame vision on a similar footing; it is not a vision-only corpus, and the image count should not be read as an ImageNet-scale advantage. From $\geq$C1.0 to $\geq$M1.0 the TSS remains of order $0.49$--$0.70$, whereas precision and F1 fall and the false-alarm rate exceeds $0.78$, so a single-frame score is a poor operational alert for rarer classes.

Sequence and video models were not trained here, but they are a natural match to the same statistics: the within-AR autocorrelation that invalidates i.i.d.\ frame counts is the temporal structure those models are meant to use, and the release already provides tracked 12~min crops, seven forecast horizons, and timestamp-aligned PNG, CSV, and MP4 records. The MP4 files are a 60~fps playback of those native frames, not a high-cadence video stream, so video architectures should be treated as temporal models on the HMI cadence rather than as conventional video recognizers. In all of these settings the independent unit remains the active region.

Analyses that assume independent and identically distributed observations, that require cycle-to-cycle hold-out, that treat MP4 clips as independent videos, or that target the rarest X-class events with fine-grained uncertainty from frame counts alone are less well supported by the nominal sample size. JW-FD is particularly suitable for AR-level generalization studies, multi-horizon and multi-threshold benchmarking, snapshot parameter or vision baselines, and sequence or multimodal experiments that exploit the aligned records. For rare-class tasks, we recommend reporting skill scores that are robust to imbalance (TSS, F1, MCC, together with precision or FAR), quantifying uncertainty at the active-region level, and, where appropriate, constructing chronological splits as noted in Section~\ref{sec:limitations}.

Beyond offline forecasting benchmarks, JW-FD can also support AI driven solar observing systems that need flare aware context or multimodal solar representations~\cite{Lin2026sidest,Tong2026astclaw,Tong2026aims}.

To reproduce or extend the dataset, users can clone the open-source pipeline at \url{https://github.com/Xiaoxuan-1/JW-FD} and set \texttt{START\_YEAR=2011}, \texttt{END\_YEAR=2025} in \texttt{config.py}. We recommend using the distributed split files so that evaluations remain comparable across publications. \del{Those files implement the 19-AR rebalance of the official split; models trained on the seed-62 membership cannot be evaluated on the released test set.}\add{Those files implement the seed~3970 assignment of Table~\ref{tab:split}. The compromise PNG working point is $B_{\mathrm{th}}=800$~G; all ten clips remain available for re-ablation.}

\subsection{Limitations}
\label{sec:limitations}

JW-FD is restricted to HMI line-of-sight magnetograms and does not ingest vector-field products such as SHARP~\cite{Bobra2014}. \del{The source full-disk frames include limb-darkening correction, but no}\add{No} geometric projection correction is applied: the released $600\times600$ crops remain in the observer plane, and $B_{\mathrm{LoS}}$ is not converted to an approximate radial field. Foreshortening and the viewing-angle dependence of the LoS component therefore remain in both the images and the 29 derived features. The $|\mathrm{longitude}| \le 60^\circ$ filter introduces a disk-center bias and excludes high-latitude and near-limb active regions; it mitigates the most severe limb effects but is not a substitute for heliographic or cylindrical equal-area remapping. AR identification depends on NOAA numbering and SRS daily positions, which can lag rapid flux emergence. The native HMI cadence of approximately 12~min may also undersample rapidly evolving pre-flare configurations.

In addition, the recommended 8:1:1 split is a stratified rejection sample of active regions across the full 2011--2025 interval\del{, with a subsequent 19-AR adjustment that balances rare-class positives between validation and test}. It therefore tests generalization to unseen active regions under similar solar-cycle conditions, and it should not be interpreted as demonstrating generalization across solar cycles or to independent future epochs. The present coverage spans the declining phase of Cycle~24 through the rising and maximum phase of Cycle~25 and cannot by itself establish cycle-to-cycle generalization. Users interested in temporal generalization may construct chronological evaluation splits from the released timestamps.

\section{Conclusions}

JW-FD is a 15-year multimodal dataset for solar-flare forecasting. It comprises 3{,}083 independent NOAA active regions and 2.00 million coregistered line-of-sight magnetogram crops spanning 1~January~2011 through 31~December~2025, released as aligned FITS, PNG, CSV, and MP4 products. PNG images, 29-dimensional magnetic features, and flare labels are provided at ten saturation thresholds from 200 to 2000~G, together with an AR-level 8:1:1 split \chg{that limits temporal leakage and, after a 19-AR rare-class adjustment, matches validation and test positive-frame counts at all seven forecast horizons}{that limits temporal leakage and, by stratified rejection sampling (seed~3970), matches rare-class positives between validation and test at all seven forecast horizons}. \add{The compromise PNG working point is $B_{\mathrm{th}}=800$~G.} JW-FD therefore supplies a large, leakage-controlled benchmark that links HMI magnetic observations with NOAA active-region and flare information across multiple prediction timescales and flare thresholds.

The dataset is intended as common, reusable infrastructure for traditional machine learning, deep learning, and multimodal methods. The baseline experiments show that the released records can support both parameter-based and vision-based models. Subsequent work can build on these records by exploiting the evolution videos, fusing image and parameter branches, incorporating vector magnetic-field products, and constructing chronological evaluation splits when cycle-to-cycle generalization is the scientific goal. We regard the present release as a foundation for that future investigation and optimization.

\authorcontributions{Conceptualization, S.M. and J.L.; methodology, S.M.; software, S.M. and L.T.; validation, S.M., H.W. and Y.L.; data curation, S.M.; writing---original draft preparation, S.M.; writing---review and editing, S.M., J.L., H.W., L.T., C.Y., Y.Z. and Y.L.; visualization, S.M.;  project administration, S.M. All authors have read and agreed to the published version of the manuscript.}

\funding{This research was supported by the National Astronomical Observatories, Chinese Academy of Sciences (No.~E4TQ2101), with additional support from the Specialized Research Fund for the State Key Laboratory of Solar Activity and Space Weather.}

\institutionalreview{Not applicable.}

\informedconsent{Not applicable.}

\dataavailability{The JW-FD construction pipeline is openly available at \url{https://github.com/Xiaoxuan-1/JW-FD}. Zenodo version~v1 (\url{https://doi.org/10.5281/zenodo.21672850}) is a representative subset of about 10\% of an earlier release. Version~v2 (\url{https://doi.org/10.57760/sciencedb.019mg}), the release corresponding to this paper, is deposited in the Science Data Bank (ScienceDB). Relative to the published JW-FD~v1, this version restores the many flares that the earlier catalog omitted by retaining only events marked with a plus sign, and rebuilds the training, validation, and test partitions so that their sample distributions are balanced. The number of frames with an M1.0 or stronger flare within the next 24~h increases from 74{,}000 in v1 to 103{,}626 in v2. September 2011 magnetograms that an earlier build did not read are included in this count.}

\acknowledgments{Open SDO/HMI magnetograms and NOAA/SWPC flare products made this release possible; we gratefully acknowledge both teams. Support from the National Astronomical Data Center (NADC) is also acknowledged.}

\conflictsofinterest{The authors declare no conflicts of interest.}

\begin{adjustwidth}{-\extralength}{0cm}
\bibliography{references}
\end{adjustwidth}

\end{document}

%% file: figures/label_timeline.tex
\begin{figure}[H]
\centering
\small
\begin{tikzpicture}[
  x=1.0cm, y=0.65cm,
  lbl/.style={font=\scriptsize},
  arr/.style={-{Stealth[length=2mm]}, semithick}
]
  \draw[thick,->] (0,0) -- (12.2,0) node[right, lbl] {time};

  \foreach \x/\t in {0/{$t_0$}, 4/{$\mathrm{Begin}-h$}, 8/{$\mathrm{Begin}$}, 11/{$t_1$}} {
    \draw (\x,0.07) -- (\x,-0.07);
    \node[lbl, below=3pt] at (\x,0) {\t};
  }

  \draw[dashed, gray!60] (8,0.12) -- (8,1.10);

  \fill[blue!12] (4,0.18) rectangle (8,1.05);
  \node[lbl, blue!60!black, align=center] at (6,0.62)
    {positive window\\$[\mathrm{Begin}-h,\,\mathrm{Begin})$};

  \fill[red!70!black] (2.0,0) circle (1.7pt);
  \fill[green!50!black] (6.0,0) circle (1.7pt);
  \fill[orange!70!black] (10.0,0) circle (1.7pt);

  \draw[arr, red!70!black] (2.0,1.85) -- (2.0,0.12);
  \node[lbl, red!70!black, align=center, above=1pt] at (2.0,1.85)
    {magnetogram at $t$\\label $= 0$};

  \draw[arr, green!50!black] (6.0,2.05) -- (6.0,1.08);
  \node[lbl, green!50!black, align=center, above=1pt] at (6.0,2.05)
    {magnetogram at $t$\\label $= 1$ if flare $\geq\tau$};

  \draw[arr, orange!70!black] (10.0,1.85) -- (10.0,0.12);
  \node[lbl, orange!70!black, align=center, above=1pt] at (10.0,1.85)
    {magnetogram at $t$\\label $= 0$ (after $\mathrm{Begin}$)};
\end{tikzpicture}
\caption{Pre-eruption labeling window for horizon~$h$. A magnetogram observed at time~$t$ receives a positive binary label for threshold~$\tau$ only if $t \in [\mathrm{Begin}-h,\,\mathrm{Begin})$ and the matched GOES class meets or exceeds~$\tau$. Observations at or after $\mathrm{Begin}$ (dashed line) are excluded from the positive class.}
\label{fig:label_timeline}
\end{figure}

%% file: figures/year_chart.tex
\begin{figure}[H]
\centering
\small
\begin{tikzpicture}
  \begin{axis}[
    ybar,
    bar width=8pt,
    width=0.92\textwidth,
    height=5.5cm,
    ymin=0,
    ymax=40,
    ytick={0,10,20,30},
    scaled y ticks=false,
    ylabel={Number of samples ($\times 10^{4}$)},
    xlabel={Year},
    symbolic x coords={2011,2012,2013,2014,2015,2016,2017,2018,2019,2020,2021,2022,2023,2024,2025},
    xtick=data,
    x tick label style={rotate=45, anchor=east, font=\scriptsize},
    y tick label style={
      font=\scriptsize,
      /pgf/number format/fixed,
    },
    point meta=explicit,
    nodes near coords={%
      \pgfmathprintnumber[fixed,precision=0,1000 sep={,}]{\pgfplotspointmeta}%
    },
    every node near coord/.append style={
      font=\tiny, rotate=90, anchor=west,
    },
    enlarge x limits=0.05,
    clip=false,
  ]
    \addplot[fill=blue!35, draw=blue!60] coordinates {
      (2011,15.3727) [153727]
      (2012,16.4070) [164070]
      (2013,18.5954) [185954]
      (2014,21.4503) [214503]
      (2015,14.1304) [141304]
      (2016,7.8828)  [78828]
      (2017,4.3320)  [43320]
      (2018,1.5317)  [15317]
      (2019,0.8281)  [8281]
      (2020,1.9166)  [19166]
      (2021,6.4611)  [64611]
      (2022,15.8002) [158002]
      (2023,24.1428) [241428]
      (2024,27.4395) [274395]
      (2025,24.1063) [241063]
    };
  \end{axis}
\end{tikzpicture}
\caption{Annual distribution of magnetogram samples in JW-FD. Sample counts track solar activity, with peaks near the Cycle~24 maximum (2013--2014) and the Cycle~25 rise (2023--2025), and a clear minimum in 2018--2020.}
\label{fig:year_chart}
\end{figure}